%% file: main.tex
\documentclass[runningheads]{llncs}

\usepackage[T1]{fontenc}
\usepackage{url}
\usepackage{hyperref}

\usepackage{multirow}
\usepackage{tableof}

\usepackage{xcolor}
\usepackage{amsmath,amsfonts}
\usepackage{graphicx}
\usepackage{xspace}
\usepackage{hhline}
\usepackage{makecell}
\usepackage{colortbl} 

\usepackage{amssymb}

\usepackage{algorithm}
\usepackage[noend]{algpseudocode}
\usepackage[inline]{enumitem}

\usepackage{orcidlink}
\usepackage[misc,geometry]{ifsym}

\usepackage[final,nomargin,inline,index]{fixme} 

\usepackage{subcaption}

\input{macros}

\title{Boosting MCSat Modulo Nonlinear Integer Arithmetic via Local Search}

\author{Enrico Lipparini\inst{1,2}\textsuperscript{(\href{mailto:enrico.lipparini@unica.it}{\Letter})}\orcidlink{0009-0009-0428-4403}\and
  Thomas Hader\inst{3}\and
  Ahmed Irfan\inst{4}\orcidlink{0000-0001-7791-9021} \and \\
  St\'ephane Graham-Lengrand\inst{4}\orcidlink{0000-0002-2112-7284} 
}
\authorrunning{Lipparini et al.}%

\institute{
  University of Cagliari, Cagliari, Italy\\
  \and
  University of Genoa, Genoa, Italy 
  \and
  TU Wien, Vienna, Austria \\
  \and
  SRI International, Menlo Park, CA, USA \\
}

\newcommand{\nconflictsthreshold}{50\xspace}

\begin{document}

\maketitle

\setcounter{footnote}{0} 

\begin{abstract}
\input{abstract}
\end{abstract}

\input{intro}

\input{background}

\input{method}

\input{localsearchNIA}

\input{experiments}

\input{relwork}

\input{conclusion}

\begin{credits}
\subsubsection{\ackname}
This material is based upon work supported in part by NSF grant 2016597. 
Any opinions, findings and conclusions or recommendations expressed in this material are those of the author(s) and do not necessarily reflect the views of the US Government or NSF.

\noindent
The work of Enrico Lipparini was partially supported by project PRIN 2022 DeLiCE (F53D23009130001) under the MUR
National Recovery and Resilience Plan funded by the European Union – NextGenerationEU.

	\subsubsection{\discintname}
	The authors have no competing interests to declare that are relevant to the content of this article.
\end{credits}

\bibliographystyle{plain}
\bibliography{refs}


\end{document}

%% file: macros.tex
\fxusetheme{color}
\FXRegisterAuthor{enrico}{anenrico}{\color{red} {\underline{enrico}}}
\FXRegisterAuthor{ahmed}{anahmed}{\color{magenta} {\underline{ahmed}}}
\FXRegisterAuthor{thomas}{anthomas}{\color{blue} {\underline{thomas}}}
\FXRegisterAuthor{stephane}{anstephane}{\color{orange} {\underline{stephane}}}

\newcommand{\ie}{i.e.\xspace}
\newcommand{\eg}{e.g.\xspace}
\newcommand{\wrt}{w.r.t.\xspace}

\newcommand{\mcsat}{MCSat\xspace}
\newcommand{\cdclt}{CDCL(T)\xspace}

\newcommand{\sota}{state-of-the-art\xspace}

\newcommand{\smt}{SMT\xspace}

\newcommand{\smtlib}{SMT-LIB\xspace}

\newcommand{\lia}{\ensuremath{\mathcal{LIA}}\xspace}
\newcommand{\nia}{\ensuremath{\mathcal{NIA}}\xspace}
\newcommand{\nra}{\ensuremath{\mathcal{NRA}}\xspace}
\newcommand{\nta}{\ensuremath{\mathcal{N\mkern-1mu T\mkern-4mu A}}\xspace}

\newcommand{\yices}{\textsc{Yices2}\xspace}

\newcommand{\yicesBase}{\ensuremath{\yices_{base}}\xspace}
\newcommand{\yicesLS}{\ensuremath{\yices_{LS}}\xspace}
\newcommand{\yicesOpt}{\yicesLS}

\newcommand{\portfolioZY}{\ensuremath{\textsc{Portfolio}}\xspace}

\newcommand{\mathsat}{\textsc{MathSAT5}\xspace}
\newcommand{\cvctool}{\textsc{cvc5}\xspace}

\newcommand{\hybridsmt}{\textsc{HybridSMT}\xspace}
\newcommand{\zthree}{\textsc{Z3}\xspace}

\newcommand{\portfolioCellZ}{\makecell{\makecell{\portfolioZY \\ (\zthree+\\\yicesLS)}\\  }}
\newcommand{\portfolioCellHybrid}{\makecell{\makecell{\portfolioZY \\ (\hybridsmt\\+\yicesLS)}\\  }}

\newcommand{\Vars}{\ensuremath{\mathit{Vars}}\xspace}

\newcommand{\VarsPhi}{\ensuremath{\Vars({\phi})}\xspace}

\newcommand{\valueM}[2]{\ensuremath{v[#1](#2)}}

\newcommand{\undefValue}{\ensuremath{\mathtt{undef}}\xspace}

\newcommand{\fixed}{\ensuremath{\mathit{VarsFixed}}}

\newcommand{\ITE}{\ensuremath{ITE}}

\newcommand{\LtoO}{\ensuremath{\mathcal{L}2\mathcal{O}}\xspace}
\newcommand{\defas}{\ensuremath{\stackrel{\text{\tiny def}}{=}}\xspace}

\ifx\hideTODO \undefined

\newcommand\ELtodo[1]{\todo[backgroundcolor=yellow,]{#1}}

\else

\newcommand\ELtodo[1]{}
\fi

\newcommand{\distLtoOterm}[2]{\ensuremath{\distLtoOtermNoargs({#1},{#2})}}
\newcommand{\distLtoOtermNoargs}{\ensuremath{d}}

\newcommand{\distLtoOnoargs}{\ensuremath{\mathrm{d}}\xspace}

\newcommand{\LS}{\ensuremath{\textproc{LS}}\xspace}

\newcommand{\trail}{\ensuremath{M}\xspace}
\newcommand{\cache}{\ensuremath{cache}\xspace}

\newcommand{\true}{\texttt{true}\xspace}
\newcommand{\false}{\texttt{false}\xspace}

\algnewcommand{\algorithmicand}{\textbf{ and }}
\algnewcommand{\algorithmicor}{\textbf{ or }}
\algnewcommand{\algorithmicnot}{\textbf{ not }}
\algnewcommand{\algorithmibreak}{\textbf{break}}
\algnewcommand{\algorithmicontinue}{\textbf{continue}}

\algnewcommand{\OR}{\algorithmicor}
\algnewcommand{\AND}{\algorithmicand}
\algnewcommand{\NOT}{\algorithmicnot}
\algnewcommand{\Break}{\algorithmibreak}
\algnewcommand{\Continue}{\algorithmicontinue}

\newcommand{\costfx}{\ensuremath{f_c}\xspace}

\newcommand{\varlist}{\ensuremath{\mathit{vars}}\xspace}

\newcommand{\nvarvisited}{\ensuremath{\mathit{n\_vars}}\xspace}
\newcommand{\currvar}{\ensuremath{\mathit{x}}\xspace}
\newcommand{\currval}{\ensuremath{\mathit{\alpha}}\xspace}
\newcommand{\newval}{\ensuremath{\mathit{\alpha_{new}}}\xspace}
\newcommand{\bestassgn}{\assgnls}
\newcommand{\bestcost}{\ensuremath{\mathit{cost^*}}\xspace}
\newcommand{\newassgn}{\ensuremath{\assgn_{new}}\xspace}
\newcommand{\newcost}{\ensuremath{\mathit{cost_{new}}}\xspace}

\newcommand{\improved}{\ensuremath{\mathit{success}}\xspace}

\newcommand{\assgn}[1][]{%
	\ifthenelse{\equal{#1}{}}{\ensuremath{\mu}\xspace}{\ensuremath{\mu_{#1}}\xspace}%
}
\newcommand{\assgnls}{\ensuremath{\assgn^*}\xspace}
\newcommand{\assgninit}{\ensuremath{\assgn[0]}\xspace}

\newcommand{\length}{\ensuremath{\mathrm{len}}\xspace}
\newcommand{\feasible}{\ensuremath{\mathit{feasible}}\xspace}
\newcommand{\feasibleM}{\ensuremath{\feasible_M}\xspace}
\newcommand{\pickfeasible}{\ensuremath{\mathit{pick\_value}}\xspace}

\newcommand{\movetofront}{\ensuremath{\mathrm{to\_front}}\xspace}

\newcommand{\movefx}{\ensuremath{\textproc{Move}}\xspace}
\newcommand{\movesRel}{\ensuremath{moves}\xspace}

\newcommand{\movepick}{\ensuremath{\movefx.\text{choose}}\xspace}
\newcommand{\movesuccess}{\ensuremath{\movefx.\text{notify}}\xspace}

\newcommand{\mode}{\ensuremath{mode}\xspace}
\newcommand{\modebool}{\text{bool-flips}\xspace}
\newcommand{\modefs}{\text{fs-jumps}\xspace}
\newcommand{\modehill}{\text{hill-climb}\xspace}

\newcommand{\stepsize}{\ensuremath{\mathit{step\_size}}\xspace}
\newcommand{\acceleration}{\ensuremath{\mathit{acc}}\xspace}

\newcommand{\phiLS}{\ensuremath{\phi_{LS}}\xspace}
\newcommand{\phiFLit}{\ensuremath{\phi_{\bot lit}}\xspace}

\newcommand{\lit}{\ensuremath{L}\xspace}
\newcommand{\clause}{\ensuremath{C}\xspace}
\newcommand{\clauseLS}{\ensuremath{C_{LS}}\xspace}

\newcommand{\leftI}{\ensuremath{\text{left}}\xspace}
\newcommand{\rightI}{\ensuremath{\text{right}}\xspace}

\usepackage{cleveref}
\usepackage{varwidth}

\makeatletter
\let\OldStatex\Statex
\renewcommand{\Statex}[1][3]{%
	\setlength\@tempdima{\algorithmicindent}%
	\OldStatex\hskip\dimexpr#1\@tempdima\relax}

\newcommand{\bestTot}[1]{\textbf{\underline{#1}}}
\newcommand{\bestPartial}[1]{\underline{#1}}

\definecolor{mygray}{gray}{0.35} 
\newcommand{\stylePortfolio}[1]{\textcolor{mygray}{{#1}}}

\newcommand{\lscalls}{\text{ls\_calls}\xspace}



%% file: abstract.tex
	%
	%
	%
	The Model Constructing Satisfiability (\mcsat) approach to the SMT problem extends the ideas of CDCL from the SAT level to the theory level. 
	Like SAT, its search is driven by incrementally constructing a model by assigning concrete values to theory variables and performing theory-level reasoning to learn lemmas when conflicts arise. 
	Therefore, the selection of values can significantly impact the search process and the solver's performance. 
	In this work, we propose guiding the MCSat search by utilizing assignment values discovered through local search. 
	First, we present a theory-agnostic framework to seamlessly integrate local search techniques within the MCSat framework. 
	Then, we highlight how to use the framework to design a search procedure for (quantifier-free) Nonlinear Integer Arithmetic (\nia), utilizing accelerated hill-climbing and a new operation called \emph{feasible-sets jumping}.
	We implement the proposed approach in the \mcsat engine of the \yices solver, and empirically evaluate its performance over the \nia benchmarks of \smtlib.


%% file: intro.tex

\section{Introduction}

Satisfiability Modulo Theory (\smt) is the problem of deciding the satisfiability of a first-order formula with respect to defined background theories. 
\smt solvers are the core backbone of a vast range of verification and synthesis tools that require reasoning about expressive logical theories such as real/integer arithmetic~\cite{smt,DBLP:journals/cacm/MouraB11}. 
One of the major \sota approaches to SMT is the Model Constructing Satisfiability calculus (\mcsat)~\cite{mcsat2,mcsat}, 
which generalizes the ideas of
  CDCL to the theory level, and which has been shown to perform particularly well on complex theories such as nonlinear arithmetic.
In the \mcsat approach, the solver  progressively constructs a \emph{theory model}, 
 similarly to how SAT solvers construct Boolean models.
Theory reasoning is used to assess the consistency of partial models,  provide explanations of infeasibility, decide theory variables, and propagate theory constraints.

When extending the partial model with a new assignment to a theory variable, picking a good value is critical for the overall performance of the solver. 
Heuristics used by state-of-the-art solvers pick values on the basis  of compatibility with the current search state  and  of computational cheapness.
This has a major drawback: 
these heuristics only consider knowledge of the current search state, neglecting information on how likely a particular assignment is to lead to a satisfying model eventually.
%

In this work, we address the problem of choosing good values for variable decisions by augmenting the  current search state knowledge 
with insights provided by local search techniques.
Following the logic-to-optimization approach~\cite{xsat,ATVApaper}, we associate to the logical formula  a cost function that represents the \emph{distance from a model}, 
 and use
local search  
 to find assignments that have a small cost. 
These assignments are then used to guide future \mcsat decisions.
%

Although local search has already been used in the context of SMT, either as a standalone solver~\cite{EfficientLocalSearch,LocalSearch4SatPolFormulas,LocalSearchIA} or as a CDCL(T) theory solver~\cite{localsearchCDCLT}, 
our work is the first to propose a tight integration of local search within the \mcsat framework,
creating a powerful synergy between the reasoning capabilities of \mcsat and the intuition provided by local search
which boosts performance for both satisfiable and unsatisfiable instances.
Our novel approach is flexible enough to allow calls to
local search at any point during the \mcsat search, seamlessly fitting with the current state. 
As \mcsat progresses through decisions, propagations, and conflicts, 
the local search problem is instantiated accordingly:
\begin{enumerate*}[label=(\roman*)]
\item 	
the cost function is built upon 
the simplification of the original formula under current state assumptions,
\item 
initial local search assignments are based on the current search state as well as cached values,
and
\item local search moves 
are enhanced by 
information on intervals of feasible assignments to theory variables as tracked by the \mcsat engine.
\end{enumerate*}

While our approach can be applied to any theory supported by \mcsat, in this work we showcase its application to the theory of nonlinear integer arithmetic (\nia).
In particular, we design a procedure based on a new operation called \emph{feasible-sets jumping}, which allows to move between feasible intervals,  and on accelerated  hill-climbing~\cite{HillClimbing}, to move inside feasible intervals.

\vspace{-0.09cm}
\paragraph{Contributions.} In this work we:
\begin{enumerate*}[label=(\roman*)]
	\item design a theory-agnostic framework to tightly integrate local search techniques within the \mcsat approach in order to guide variable decisions,
	\item use the framework to define a local search procedure for the theory of nonlinear integer arithmetic, that makes use of feasible-sets jumping and accelerated hill-climbing, and
	\item show the practical applicability of our method using our implementation in the \mcsat engine of \yices~\cite{yices} on the quantifier-free \nia benchmark set of \smtlib~\cite{SMTLIB}.
\end{enumerate*}

\vspace{-0.09cm}
\paragraph{Structure.} 
In Section~\ref{sec:background}, we provide the necessary background. 
Section~\ref{sec:mcsatLS} describes a deep integration of local search techniques within the MCSat framework from a general point of view,
which is applied in Section~\ref{sec:LSarith} to define a local search approach for non-linear integer arithmetic.
In Section~\ref{sec:Experiments}, we show and discuss the results of our experiments before presenting related work in Section~\ref{sec:relwork} and concluding in Section~\ref{sec:conclusion}.

%% file: background.tex
\section{Preliminaries}
\label{sec:background}

We assume basic knowledge on the standard first-order quantifier-free logical setting and standard
notions of theory, satisfiability, and logical consequence.
%
We write logical \emph{variables} with $x,y,\ldots$, and concrete \emph{values} 
with $\alpha, \beta, \ldots$ (the domain of concrete values is theory specific, \eg $\mathbb{Z}$  for integer arithmetic, $\mathbb{R}$ for real arithmetic).
An \emph{assignment} $\assgn$ is a map from variables to values of matching type.
If $\phi$ is a formula, we denote with $\VarsPhi$ the set of its (free) variables.
We use $C$ to denote clauses, and $L$ to denote literals.
Nonlinear Integer Arithmetic (\nia) is the theory consisting of arbitrary Boolean combinations of Boolean variables and arithmetic atoms of the form of polynomial equalities and polynomial inequalities over integer variables. 
It is undecidable by Matiyasevich’s theorem~\cite{Matiyasevich}.

%


\subsection{SMT \& MCSat} 
\label{sec:background:mcsat}

\smt~\cite{smt} is the problem of deciding
the satisfiability of a first-order formula with respect to some
theory or combination of theories.
Two of the major approaches for SMT solving are the Conflict-Driven Clause Learning with theory support (\cdclt)~\cite{dpllt,smt} and the \mcsat approach. 
In the former, theory solvers augment a propositional SAT engine with theory reasoning procedures which are capable of deciding a conjunction of literals (i.e. atomic formulas and their negations) in a particular theory.
A propositional model (of the Boolean abstraction of the formula) found by the SAT solver 
is then checked by all theory engines for theory consistency.

\begin{figure}[t]
	\centering
	\includegraphics[scale=.36,clip]{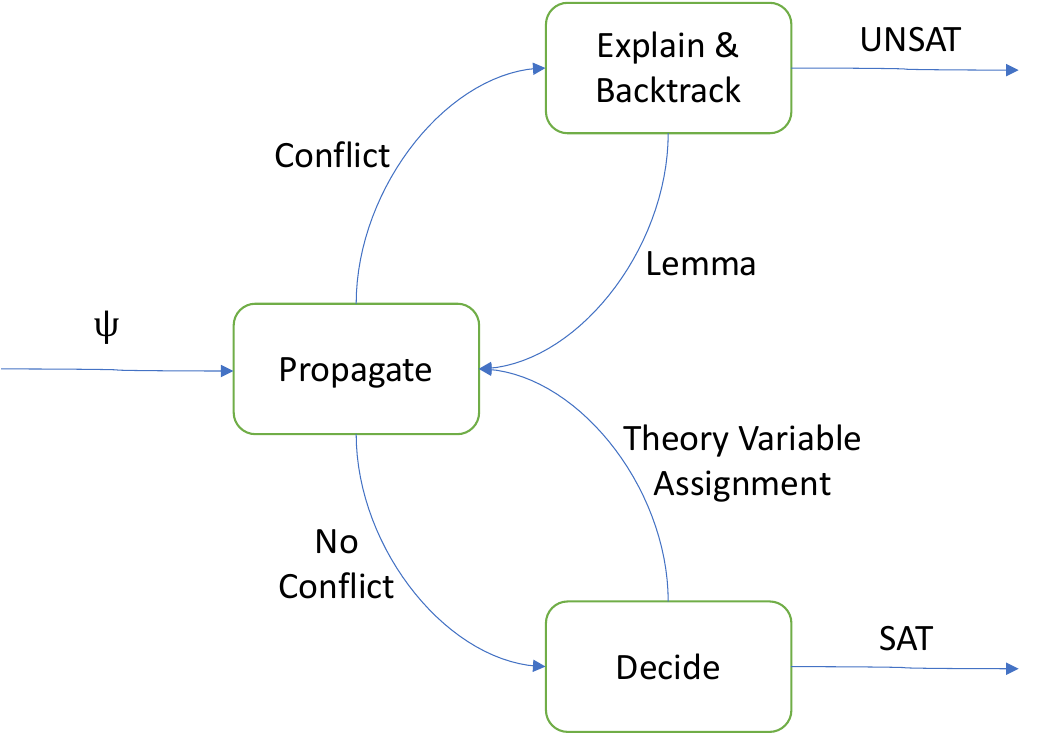}
	\caption{\small The \mcsat framework consists of the following steps:
		1) Propagate the trail.
		2) If a conflict is found during propagation, check if there is
		any decision to backtrack over. If not, return
		UNSAT. Otherwise, explain the conflict using a lemma, backtrack the
		trail, and repeat step 1.
		3) If no conflict is found during propagation, decide on a variable
		that is not on the trail. If there is nothing left to decide, return
		SAT. Otherwise, add the decided variable to the trail and repeat
		step 1.}
	\label{fig:mcsat-arch}
\end{figure}

The latter, \mcsat, applies CDCL-like mechanisms to perform theory reasoning directly.
It can be used either as a theory solver for a specific theory (e.g. in Z3~\cite{z3} for non-linear arithmetic over the reals and the integers~\cite{z3ArithSolving}), 
or as a fully-fledged stand-alone engine able to handle multiple theories 
(e.g. in Yices2 for non-linear arithmetic over the reals~\cite{solving_nonlin} and over the integers~\cite{branchBoundJovanovic}, bit-vectors~\cite{mcsatBV},  arrays~\cite{mcsatArrays}, and finite fields~\cite{mcsatFF1,mcsatFF2};
 as well as in SMT-RAT~\cite{SMTRAT} for non-linear real arithmetic~\cite{SMTRATmodular}). 
 The MCSat architecture consists of a core solver, an
assignment trail, and plugins for theory reasoning.
Figure~\ref{fig:mcsat-arch} illustrates the high level
flow of the \mcsat framework.

The core solver incrementally constructs a partial model consisting of Boolean and theory assignments (stored in a \emph{trail}),
maintaining the invariant that none of the constraints evaluate to false under the partial model.
The trail contains three kinds of elements: \emph{propagated literals} (literals implied to be true by the current state), \emph{decided literals} (literals that we assume to be true), and \emph{model assignments} (assignments of first-order variables to concrete values).
Propagations, conflict analysis, lemmas generation, and variable decisions are all handled by theory plugins (including a Boolean plugin that is responsible for propositional reasoning).
In general, plugins also keep a  \emph{feasibility set} for each variable of their competence,
containing the values that are consistent with the current trail and are, thus, candidates to be picked for deciding the variable.


When the core solver selects a variable for decision, the choice of the value to assign to the variable is handled by the theory plugin responsible for its type.
Some solvers (\eg Yices2) implement a heuristic called \emph{value cache}
(a generalization of SAT {phase saving}~\cite{pipatsrisawat2007lightweight}),
that keeps track of the last value assigned to a variable when the assignment is undone. 
Then, when a decision has to be made for the variable, the cached value will be used, provided that it is still in the feasibility set; otherwise, it will simply be ignored, and other heuristics will be used (\eg picking a default value such as $0$).

In the following, we will denote with \trail the trail, with $\valueM{\trail}{x}$ the value of the variable $x$ in the trail (which may be equal to \undefValue if the variable is not assigned in the trail), and with $\valueM{\trail}{L}$ the value of the literal $L$ under the assignment in $\trail$ which may be true ($\top$) or false ($\bot$) if $L$ can be fully evaluated under such assignment, or  \undefValue otherwise.
We denote with $\feasibleM(x)$ the feasibility set of $x$ in $\trail$, i.e.\ all values that can be chosen for $x$ in the current search state given by $\trail$.
For arithmetical theories, we have that $\feasibleM(x) \subseteq \mathbb{R}$, and, in particular, that $\feasibleM(x)$ is the union of a finite set of \emph{feasible intervals}, \ie $\feasibleM(x) = \bigcup_{i\in [0:m]}I_i$. 
We assume that theory plugins provide a function $\pickfeasible(S)$ that returns a value from a set $S$.

\begin{example}\label{ex:trail}
	\newcommand{\cA}{\ensuremath{(x \geq 1)}}
	\newcommand{\cB}{\ensuremath{(xy = 1)}}
	\newcommand{\cC}{\ensuremath{(x + 2yz > 0)}}
	\newcommand{\cD}{\ensuremath{(z^2 > 1)}}
	\newcommand{\prop}[1]{\textcolor{darkgray}{#1}}
	\newcommand{\deci}[1]{\textcolor{black}{#1}}
	\newcommand{\B}{$\mathbb{B}$\xspace}
	\newcommand{\Z}{$\mathbb{Z}$\xspace}
	Assume a search problem in \Z with variables $x$, $y$, and $z$ given by the input formula $\phi$.
	\[ \phi \;=\; (\lnot \cA \lor \cB) \;\land\; (\lnot\cB \lor \cC) \;\land\; \cD \]
	A possible trail at some point during the search is
	\[\trail = [ \prop{\cD\mapsto\top}, \deci{x \mapsto 1}, \prop{\cA \mapsto \top}, \prop{\cB\mapsto\top} ]\]
	On \trail elements are either \deci{decided} or \prop{propagated}.
	The feasibility sets are given by	
	 ${\feasibleM(z) = (-\infty, -1) \cup (1, \infty)}$ and $\feasibleM(y) = \{1\} = \feasibleM(x)$.
	 Since $\feasibleM(y)$ is a singleton,  we can propagate the assignment $y \mapsto 1$ on $\trail$.
	We further have that $\valueM{\trail}{\cC} = \undefValue$ and $\valueM{\trail}{x} = 1$.
	\let\cA\undefined
	\let\cB\undefined
	\let\cC\undefined
	\let\cD\undefined
	\let\B\undefined
	\let\Z\undefined
	\let\prop\undefined
	\let\deci\undefined
\end{example}

\subsection{Local Search}
\label{sec:background:ls}

We define a local search problem as a triple $(\assgn[0], \costfx, \movesRel)$, where:
\begin{itemize}
	\item \assgninit is an initial assignment for a set of variables \Vars,
	\item $\costfx$ is a cost function from the set of assignments to $\mathbb{R}_{\geq0}$,
	\item $\movesRel$ is a neighbor relation between assignments. 
\end{itemize}

A local search algorithm starts from the initial assignment \assgninit and iteratively explores neighboring assignments according to the \movesRel relation. 
We say that $\assgn'$ is a \emph{move from \assgn} if $(\assgn, \assgn') \in \movesRel$.
A move $\assgn'$ is \emph{accepted} if $\costfx(\mu') < \costfx(\mu)$. 
When a move is accepted, the new assignment becomes the current assignment and the search continues
until either: a zero-cost assignment is found, there are no more possible moves (meaning that the current assignment represents a \emph{local minimum}), or a given stopping criterion is reached (\eg number of moves).

\sloppy{The problem of finding a solution for an SMT formula $\phi$ can be encoded as a local search problem,
\eg,
by following the logic-to-optimization approach~\cite{xsat,ATVApaper,Lipparini:23}, 
in which
a formula $\phi$ is mapped to a 
term
$\LtoO(\phi)$ that represents the \emph{distance from a solution}. }

In principle, the \LtoO operator can be defined for any theory for which the concept of distance between terms makes sense. Here, we limit ourselves to arithmetic theories.
%
%
%
%
We introduce an arithmetic function symbol $d$ of arity 2 and we assume a fixed interpretation $\distLtoOnoargs$ 
that satisfies the properties of metric distance, i.e. symmetry, positivity, reflexivity, and triangle inequality.
We also assume the existence of a 
fixed constant term $\epsilon$, such that $\epsilon > 0$.
The specific choice of $\distLtoOnoargs$ and $\epsilon$  is theory-dependent. 

We recursively define \LtoO as follows:
\begin{displaymath}
	\begin{array}{lll}
		
		\LtoO(b) &\defas& \ITE(b, 0, 1) \\
		\LtoO(\neg b) &\defas&  \ITE(b, 1, 0) \\
		\LtoO(t_1 = t_2) &\defas& \distLtoOterm{t_1}{t_2} \\
		\LtoO(t_1 \leq t_2) &\defas& \ITE(t_1 \leq t_2, \ 0 , \ \distLtoOterm{t_1}{t_2})  \\
		\LtoO(t_1 < t_2) &\defas& \ITE(t_1 < t_2, \ 0 , \ \distLtoOterm{t_1}{t_2} + \epsilon)  \\
		\LtoO(t_1 \neq t_2) &\defas& \ITE(t_1 \neq t_2, \ 0 , \ 1)  \\
		\LtoO(\phi_1 \land \phi_2) &\defas& \LtoO(\phi_1) + \LtoO(\phi_2) \\
		\LtoO(\phi_1 \lor \phi_2) &\defas& \LtoO(\phi_1)  \cdot \LtoO(\phi_2)
		\\
		\LtoO(ITE(\phi_{c}, \phi_1, \phi_2) &\defas&  \ITE(\phi_c,\ \LtoO(\phi_1),\ \LtoO(\phi_2)) \\
		\LtoO(\neg \ITE(\phi_{c}, \phi_1, \phi_2)) &\defas&  \ITE(\phi_c,\ \LtoO(\neg \phi_1),\ \LtoO(\neg \phi_2)) \\
	\end{array}
\end{displaymath}

It is easy to check that a complete assignment $\assgn$ satisfies $\phi$ if and only if 
$\LtoO(\phi)$ evaluates to $0$ under $\mu$.
 
In the following, with a slight abuse of notation, we denote with $\LtoO(\phi)$ also the corresponding arithmetic function determined by the interpretation $\distLtoOnoargs$ and the constant $\epsilon$, and
we define the \emph{cost function}  associated to $\phi$ as $\costfx \defas \LtoO(\phi)$.

\begin{example}
Let $\phi \defas b \ \land \ x = y^2$, and $\distLtoOnoargs(t_1, t_2) \defas |t_1 - t_2|$. 
Then, the cost function associated to $\phi$ is 
$${\costfx \defas \LtoO(\phi) = \ITE(b, 0, 1) + |x-y^2|}$$ 
Now, let  
$\assgn[0] \defas \{b\mapsto \bot\ ;\ x \mapsto 4 \ ;\ y \mapsto 1\}$ be a starting assignment.
We have that 
$\costfx(\assgn[0]) = \ITE(\bot, 0, 1) + |4-1^2| = 1 + 3 = 4$.
If we consider the move ${\assgn[1] \defas \{b\mapsto \top\ ;\ x \mapsto 4 \ ;\ y \mapsto 1\}}$ that flips $b$,
then
$\costfx(\assgn_1) = \ITE(\top, 0, 1) + {|4-1^2| = 0 + 3 =3}$, hence the move is improving and is accepted.
Then, if we consider the move $\assgn[2] \defas \{b\mapsto \top\ ;\ x \mapsto 4 \ ;\ y \mapsto 2\}$ that increases the value of $y$ by $1$, 
we have 
$\costfx(\assgn_1) = \ITE(\top,0, 1) + |4-2^2| = 0 + 0 = 0$. Hence we have found a zero for $\costfx$, \ie a satisfying assignment for $\phi$.
\end{example}

In general, local search is not guaranteed to find a solution of $\phi$, if there is any.
Nevertheless, it returns a \emph{local minimum}/\emph{best-effort value} of the cost function in the neighborhood of the initial assignment.

%% file: method.tex

\section{Deep combination of Local Search and MCSat}

\label{sec:mcsatLS}


We propose a deep combination of MCSat and local search 
where:
\begin{enumerate*}[label=(\roman*)]
\item the current  state  of MCSat is used to instantiate a local search problem
and \item the results of the local search help guiding future MCSat decisions. 
\end{enumerate*}
%
Assuming we have a local search procedure \LS,
we discuss how to instantiate \LS (\Cref{sec:mcsatLS:instantiateLS}),
as well as how to use the result of \LS within MCSat (\Cref{sec:mcsatLS:guideMCSat}).  

\subsection{Instantiating the Local Search problem}
\label{sec:mcsatLS:instantiateLS}

For the instantiation of \LS, we determine the initial assignment and the formula upon which the cost function is constructed.
Both choices are of fundamental importance. 
A good initial assignment is essential to find a good local minimum of the cost function. A \emph{good local minimum} is a local minimum that meets two conditions:
\begin{enumerate*}[label=(\roman*)]
\item it has a smaller cost compared to the cost of the initial assignment and 
\item its assignment values are likely to be accepted by MCSat, \ie,  they are consistent with the current trail.
\end{enumerate*}
Passing a simplified formula that takes
the truth value of propagated and decided literals into account is also essential
to tailor the search to the current \mcsat state and
to avoid unnecessary computations.

\input{alg_initLS}
\paragraph{Initial assignment.}
For every model assignment $x \mapsto \alpha$ in \trail, 
the assigned variable is treated as a constant that takes its respective assigned value
(\ie, $x$ is treated as the constant $\alpha$) and is not allowed to be changed in \LS. 
This reduces the dimension of the \LS search space, and avoids moves inconsistent with the current trail. 
For initial assignment of variables that are unassigned in \trail,
a reasonable choice is to use cached values of previous search states, if present in the value cache.
However, cached values are not guaranteed to be in the feasibility set, as they might be the result of a previous decision that eventually led to a conflict. 
Hence, we first check if the cached value is feasible.
If it is not, or there is no cached value,
we pick any value from the  feasibility set by asking the appropriate theory plugin.
The procedure for choosing the initial assignment is shown in \Cref{alg:initAssLS}.
Note that, for all the variables, the feasibility set cannot be empty. An empty feasibility set indicates an inconsistent trail which is resolved using conflict resolution before starting \LS.
\mcsat maintains the invariant that, for a consistent trail, the feasibility set of all variables is non-empty.

\input{alg_formulaLS}

\paragraph{Formula for LS.} Every Boolean assignment $\lit \mapsto \{\top, \bot\}$ in $\trail$ represents the truth value of the literal $\lit$ that is assumed to hold at the current search state (either because of a propagation or a decision).
We can use this information to simplify the original formula before passing it to  \LS. 
For a given clause 
$\clause \defas \lit \lor \lit_1 \lor \dots$, 
if $\valueM{\trail}{\lit} = \top$, 
then, for \LS, it suffices to find an assignment that satisfies $\lit$,
since such an assignment would satisfy $\clause$ as well. 
Hence, in this case, we shall pass to \LS just $\lit$ instead of $\clause$.
On the other hand, 
if $\valueM{\trail}{\lit} = \bot$,  
then there is no incentive for \LS to try to find an assignment that makes $\lit$ true, as any such assignment would be inconsistent with the trail and will be discarded by \mcsat immediately.
Thus, $\lit$ is removed from the clause that is passed to \LS. 
Note that, by just removing $\lit$ from the clause, we still may end up with an assignment that evaluates $\lit$ to $\top$.
Therefore, for literals that are assigned to $\bot$ in the trail, we add, just once, $\neg \lit$ to the formula that we pass to \LS.
This procedure is shown in \Cref{alg:formulaLS}.

\subsection{Guiding MCSat decisions} 
\label{sec:mcsatLS:guideMCSat}
During the search, we periodically call \LS to suggest values for \mcsat to choose in subsequent decisions.
As a heuristic to decide when to call \LS, we are utilizing a polynomially increasing conflict threshold. Initially, this limit is set to \nconflictsthreshold, and then it is increased according to the polynomial $50 \cdot \lscalls \cdot \log_{10}(\lscalls+9)^3$, where \lscalls represents the number of times LS has been called. 
A similar heuristic is used by SAT solvers to decide when to perform certain cache clearing operations~\cite{satCdclLStargetphase}.
Once the threshold is reached, we wait until the last conflict has been resolved and all consequences of that conflict are propagated.
Then we start \LS to guide any further decisions.

The return of \LS consists of a complete assignment 
that contains \emph{suggested values} for future variable decisions.
These suggested values are put in the MCSat value cache -- recall that the values in the cache are picked first during variable decisions, provided that they are feasible.
Note that, during the choice of the initial assignment to pass to \LS, we had relied on (feasible) cached values as well.
If a cached value was feasible and \LS changed its value, it means that such change  led to a smaller cost,
hence got us closer to a solution.
Therefore, replacing the old cached value with the newly found suggestion improves the cache quality.
On the other hand, if the cached value was not feasible, then any change to a feasible value improves the cache quality.



Furthermore, \LS keeps track of the activity of each variable during its execution.
The most active variables have contributed most to the decrease of the cost during LS. 
We suggest those variables to \mcsat as good choices for subsequent decisions.
This way,  variables that were  more active during the local search phase, 
will have a higher impact on the \mcsat search.

%% file: alg_initLS.tex
\begin{algorithm}[t!]
	\caption{Initial assignment for \LS}\label{alg:initAssLS}
	\begin{algorithmic}[1]
		\Require a set \VarsPhi, a trail \trail, a value cache \cache, a feasibility map \feasibleM
		\Ensure an initial assignment $\assgn[0]$. 
		a set of fixed variables $\fixed \subseteq \VarsPhi$ 
		\State $\fixed \gets \emptyset$
		\For{$x \in \VarsPhi$}
		\If{ $ \valueM{\trail}{x} \neq \undefValue$ } \Comment{check if $x$ has a value in the trail}
		\State $\assgn[0](x) \gets  \valueM{\trail}{x} $  \Comment{assign trail value}
		\State \fixed.add($x$)
		\ElsIf{$\cache(x) \neq \undefValue$  $\AND \cache(x)\in \feasibleM(x)$} 
		\State $\assgn[0](x) \gets  \cache(x) $ \Comment{assign feasible cached value}
		\Else
		\State $\assgn[0](x) \gets \pickfeasible(\feasibleM(x)) $  \Comment{assign any feasible value}
		\EndIf
		\EndFor
	\end{algorithmic}
\end{algorithm}

%% file: alg_formulaLS.tex
\begin{algorithm}[b!]
	\caption{Formula for \LS}\label{alg:formulaLS}
	\begin{algorithmic}[1]
		\Require a formula $\phi$, a trail $\trail$
		\Ensure a subformula $\phiLS$ of $\phi$
		\State $\phiLS \gets \top$ \Comment{formula to be passed to \LS}
		\For{$\clause \in \phi$} 
		\State $\clauseLS \gets \bot$
		\For{$\lit \in C$} 
		\If{$\valueM{\trail}{\lit} = \top$} \Comment{if literal is assigned to true in  trail}
		\State $\clauseLS \gets \top$ \Comment{substitute the clause with true}
		\State $\phiLS\gets \phiLS\land  \lit$ \Comment{store  literal}
		\State \Break
		\ElsIf{$\valueM{\trail}{\lit} = \bot$} \Comment{if literal is assigned to false in  trail}
		\State $\phiLS\gets \phiLS\land \neg \lit$\Comment{store   literal (with correct polarity)}
		\State \Continue \Comment{ignore  literal in the clause}
		\Else 
		\State $\clauseLS \gets \clauseLS \lor L$ \Comment{keep literal in the clause}
		\EndIf
		\EndFor
		\State $\phiLS \gets \phiLS \land \clauseLS$ \Comment{store simplified clause}
		\EndFor
	\end{algorithmic}
\end{algorithm}
%
%
%
%

%% file: localsearchNIA.tex
\section{Local Search for Nonlinear Integer Arithmetic}
\label{sec:LSarith}

As explained in \Cref{sec:mcsatLS:instantiateLS}, 
\LS  receives an initial assignment $\assgninit$ and a formula $\phi$ from MCSat. The formula $\phi$ is used to construct the cost function $\costfx$ using the logic-to-optimization approach (\Cref{sec:background:ls}), i.e. $\costfx \defas \LtoO(\phi)$.
To apply that, we must first
define the distance function \distLtoOnoargs and the strict inequality constant $\epsilon$ for integers. For \distLtoOnoargs, we choose a consistent and computationally cheap definition $\distLtoOnoargs(t_1, t_2) \defas |t_1 - t_2|$. 
For $\epsilon$, our choice is $\epsilon \defas 1$, since $t < 0$ is interchangeable with $t + 1 \leq 0$ for integers.

The building blocks of local search are \emph{moves}. We contemplate three types of moves (or \emph{modes}): one for Boolean variables, and two for integer variables.
Given an assignment $\assgn$, we have the following types of moves:
\begin{itemize}
	\item \emph{Boolean flips}: 
		For a Boolean variable $b$, the assignment  ${\mu_{\neg b} \defas \mu[b \mapsto \neg \mu(b)]}$ obtained by changing the value of $b$ to the negation of its value assigned by $\mu$ is a \emph{flip move} from $\mu$.
	\item \emph{Hill-climbing moves}:
		In the basic version of hill-climbing, for an integer variable $x$, the assignments 
		${\mu_{x+1} \defas \mu[x \mapsto \mu(x)+1]}$ and
		${\mu_{x-1} \defas \mu[x \mapsto \mu(x)-1]}$ obtained by mapping $x$ to the successor and predecessor of its value assigned by $\mu$  are moves from $\mu$. 
	\item \emph{Feasible-set-jumps}: 
		For an integer variable $x$, 
		with feasibility set ${\feasible(x) = \bigcup_{i\in [0:m]}I_i}$, 
		and $x\in I_j$ (for a given $j\in[0:m]$), 
		the assignments $\mu_{\leftI} \defas \mu[x \mapsto \pickfeasible(I_{j-1})]$, 
		and  $\mu_{\rightI} \defas \mu[x \mapsto \pickfeasible(I_{j+1})]$
		obtained by picking a value from the left and right feasible intervals of $I_j$ 
		(provided they exist, \ie, respectively, that $j-1 \in [0:m]$, and $j+1 \in [0:m]$) 
		are moves from $\assgn$.
\end{itemize}

\input{alg_LSmain}

Our specific strategy of the local search algorithm is outlined in \Cref{alg:LSloop}. 
The algorithm starts with a list of variables \varlist (and an associated \feasible map), an initial assignment \assgn[0], and a cost function $\costfx$.
The goal of the procedure is to return an assignment $\bestassgn$ that improves over the initial assignment \assgn[0] \wrt the cost function, \ie ${\costfx(\assgnls) < \costfx(\assgn[0])}$.

At the beginning, the best assignment coincides with the initial assignment (\Cref{alg:LSloop:bestAssInitAss}).
First, we cycle over modes (\Cref{alg:LSloop:cycleModes}).
Then, we enter in a loop over the variables (\Cref{alg:LSloop:firstLoop}). 
The loop breaks only in two cases: if all the variables have already been visited since the last improvement (in which case, it means we have reached a local minimum \wrt the current mode moves), or if the current cost is equal to $0$ (in which case it means we have found a solution).
At each loop iteration, we pick the next variable (\Cref{alg:LSloop:pickNewVar}).
Here, for simplicity, we assume that there are no fixed variables (in practice, these variables are just ignored and treated as constants).
Then, for the current variable \currvar, we enter in a second loop (\Cref{alg:LSloop:pickMove}), in which we select  new values for \currvar. 
These values are determined by a move selection module, which we discuss  below.
The loop breaks only when there are no more moves available.
For each value,
a new assignment is built by re-assigning \currvar to the new value (\Cref{alg:LSloop:newAss}), 
and the cost of the new assignment is computed (\Cref{alg:LSloop:newCost}).
We then check whether the current cost is lower than the previous cost (\Cref{alg:LSloop:improves}). 
If so, then the new assignment becomes the best assignment (\Cref{alg:LSloop:updateAss}), and \currvar is moved to the front of the list (\Cref{alg:LSloop:moveVarToFront}). 
If not, then we try other moves for \currvar, if there are any.
In both cases, we notify the move selection module whether the suggested move has led to a success or not (\Cref{alg:LSloop:notifyMove}).
The move selection module works as following. 

\paragraph{Boolean flips mode.} Here, the logic is rather straightforward, as there is only one move possible per variable.
Regardless of whether the move has success or not, the cycle over moves terminates, and the algorithm proceeds with the cycles over variables or over modes. 

\paragraph{Accelerated hill-climbing mode.} 
	The simple hill-climbing moves presented earlier, in which we add or subtract $1$ to the current value, can be quite slow in converging toward a local minimum when the search space is huge.
	For this reason, we \emph{accelerate} hill-climbing by keeping, for each variable, an adaptive \stepsize,  which is incremented or decremented according to a fixed acceleration parameter  \acceleration (in our setting $\acceleration=1.2$) and on the base of the success of previous moves.
	At the beginning, \stepsize is set to $1$ (\ie, we start with simple hill-climbing moves). 
	At each iteration, we try four moves, corresponding to adding to the current value the product between $\stepsize$ and one of the following: $\acceleration, \frac{1}{\acceleration}, \frac{-1}{\acceleration}, -\acceleration$.
	Since we are working with integers, every step value is rounded to the nearest integer.
	If one of the moves has success, then we set \stepsize to  be equal to the best successful step (thus keeping the best velocity).	
	If none of the moves has success, then we stop the moves cycle, and we set $\stepsize$ to $\frac{\stepsize}{\acceleration}$ (thus decelerating over this variable for future moves).

\paragraph{Feasible-set-jumping mode.} 
	There are two possible versions of fs-jumping: \emph{global}  and \emph{local}. 
	In global fs-jumping, given a fixed variable, we try \emph{all}  possible jumps over the feasibility set, \ie we try one jump per feasible interval. 
	While this may give 
	 a wide-ranging view over the feasibility set,
	 it can also be very costly, hence we limit global fs-jumping to one time per variable (per \LS call). 
	Local fs-jumping, on the contrary, only explores the left and the right feasible intervals \wrt to the interval that contains the current value.
	If one fs-jump is successful, \eg the one to the left interval, then we continue on that direction and explore the interval further left. 
	As soon as we find that both left and right fs-jumps do not improve, then we stop, hence avoiding to span over all feasible intervals like in the global fs-jumping.

%% file: alg_LSmain.tex
\begin{algorithm}[b!]
	\caption{\LS main algorithm}\label{alg:LSloop}
	\begin{algorithmic}[1]
		\Require a list \varlist, a feasibility map \feasible, an initial assign. \assgninit, 
		 a cost function $\costfx$
		\Ensure a final assignment $\assgnls$  with $\costfx(\assgnls) \leq \costfx(\assgninit)$
		\State $\bestassgn \gets \assgn[0]$ \Comment{best assignment} \label{alg:LSloop:bestAssInitAss}
 		\State $\bestcost \gets \costfx(\assgn[0])$ \Comment{best cost}
		\For{$\mode \in \{\modebool , \modefs, \modehill\}$} \label{alg:LSloop:cycleModes}
		\State $\nvarvisited \gets 0$ \Comment{no. of vars visited since last improvement}	
		\While{$\nvarvisited < \length(\varlist) \AND \bestcost \neq 0$}  \label{alg:LSloop:firstLoop}
			\State  $\currvar \gets \varlist[\nvarvisited]$  
			  \Comment{pick next variable} \label{alg:LSloop:pickNewVar}
			\State $\currval \gets \bestassgn(\currvar)$ \Comment{value assigned  to \currvar}
			\While{$\newval \gets \movepick(\currvar,\currval,  \feasible,  \mode)$} 
			\label{alg:LSloop:pickMove}
				\State $\newassgn \gets \bestassgn[\currvar \mapsto \newval]$ \Comment{create new assignment}  \label{alg:LSloop:newAss}
				\State $\newcost \gets \costfx(\newassgn)$  \label{alg:LSloop:newCost}
				\State $\improved \gets \newcost < \bestcost$ 	\Comment{check if the move has improved}  \label{alg:LSloop:improves}
				\If{\improved}
					\State $\bestassgn \gets \newassgn$ \Comment{update best assignment}  \label{alg:LSloop:updateAss}
					\State $\bestcost\gets \newcost$ \Comment{update best cost}  \label{alg:LSloop:updateCost}
					\State $\nvarvisited \gets 0$ \Comment{reset no. of vars visited} \label{alg:LSloop:resetNVar}
					\State $\varlist.\movetofront(\currvar)$ \Comment{move \currvar to the front of the list} \label{alg:LSloop:moveVarToFront}
				\EndIf
				\State $\movesuccess(
				\currvar, \currval, \newval, \feasible,  \mode, \improved)$ 
				 \label{alg:LSloop:notifyMove}
			\EndWhile
			\State $\nvarvisited \gets \nvarvisited +1 $ \Comment{increas no. of vars visited} \label{alg:LSloop:increaseNVar}
		\EndWhile
		\EndFor
	\end{algorithmic}
\end{algorithm}

%% file: experiments.tex
\section{Experiments}
\label{sec:Experiments}

\newcommand{\toprowtextangle}{60}
\input{results24s_vert}
\input{results300s_vert}

\paragraph{Implementation.} We have implemented our  method in
the MCSat engine of the \yices
SMT solver, 
adding a module for the interaction with LS. 
We will denote the version of \yices that makes use of \LS as \yicesLS and the baseline version (without any local search) as \yicesBase.

\paragraph{Setup.} We have run our experiments on a cluster equipped with AMD EPYC 7502 CPUs running at 2.5GHz, using a timeout of 300 seconds, and a memory limit of 8GB. 
We have compared the base version of \yices with the LS-boosted version \yicesOpt as well as with the \sota SMT solvers  \cvctool~\cite{cvc5} (version 1.2.0), \mathsat~\cite{mathsat5} (version 5.6.11), and \zthree~\cite{z3} (version 4.13.3). 
We have also included in the comparison \hybridsmt~\cite{localsearchCDCLT}, which runs a portfolio composed by the LS solver LocalSMT~\cite{LocalSearchIA} (used as a standalone tool) and a LocalSMT-boosted version of \zthree's CDCL(T)%
\ -- see also related work (\Cref{sec:relwork})%
.

\paragraph{Benchmarks.} We have considered all the SMT-LIB~\cite{SMTLIB} (Version 2024~\cite{preiner_2024_11061097}) benchmarks from the QF\_NIA category. This is a class of 25443 benchmarks, among which 14990 and 5183 come with a known status of ``sat'' and ``unsat'', respectively, and another 5270 have an ``unknown'' satisfiability status. 

\paragraph{Results.}
In the presentation of the results, we consider both a short time limit and a long time limit. 
We set the short time limit to 24s, as in the respective SMT-COMP track~\cite{SMTComp}, and the long time limit to 300s, due to resource constraints. 
The results  are shown in  \Cref{fig:res24s} and   \Cref{fig:res300s}, respectively. 
In the columns, we separate per benchmark family; on the rows, for each solver, we report the amount of overall benchmarks solved, and, in parenthesis, the amount of benchmarks solved restricted to sat and unsat instances, respectively.
We also include two portfolios between \zthree (resp., \hybridsmt) and \yicesLS, that work as follows: we run \zthree (resp., \hybridsmt) for half of the time limit (\ie, 12s/150s), then, if it has not terminated, we run \yicesLS for the remaining time.

\input{cactus_plots}

\input{scatter}

\paragraph{Discussion.}
First, we observe that, with both time limits,  \yicesOpt solves a significant number of benchmarks more than \yicesBase.
In particular, on both  satisfiable and 	unsatisfiable instances, it improves (or matches) \yicesBase results over all families, except one. 
Improving on unsatisfiable benchmarks is noteworthy:
indeed, while, in general, local search is geared toward proving satisfiability,
 integrating it within \mcsat enables to  generate better lemmas. 
 This is witnessed not only by the higher amount of benchmarks solved overall, but also by the lower amount of conflicts and theory variables assignments occurred.
 On  unsatisfiable benchmarks solved by both tools, on average (resp. median), \yicesLS encountered 225 (resp. 4) fewer conflicts and 27826 (resp. 36) fewer theory variables assignments than \yicesBase.
  Note that, on average (resp. median), \yicesBase encountered 1669 (resp. 458) conflicts and 79842 (resp. 10537) theory variable assignments. 
  These numbers show that there is a considerable amount of benchmarks for which the number of conflicts and theory variable assignments is significantly lower (note that such a lower median \wrt average implies a pronounced right-skewness).

Overall,
we see that, in the 24s track, 
\yicesLS solves more benchmarks than any other solver,
 while,
 in the 300s track,
 it comes third, after \hybridsmt and \zthree.
The complementarity of \yicesLS \wrt  both tools can be  witnessed by the scatter plots in  \Cref{fig:scatter}, and by the results of the portfolios in  \Cref{fig:res24s,fig:res300s}. 
 Note that \zthree internally utilizes portfolio tactics that combine multiple solving techniques sequentially 
 (clearly observable in \Cref{fig:cactus,fig:scatter}).
 \hybridsmt runs a higher-level portfolio that combines the LS-based LocalSMT and \zthree.
 The results of the portfolios that include \yicesLS show that our approach  brings significant diversity to the strategies already used in \sota portfolio approaches.
 %
 %
 
 Since \hybridsmt is the only other solver that -- to the best of our knowledge -- leverages local search techniques for \nia, 
 it is interesting to compare the improvements it brings to \zthree 
 with the improvements that 
 \yicesLS brings over \yicesBase.
 We can see that, with a 300s time limit, the improvements are comparable, as both tools solve around 800 benchmarks more than their base solvers.
 With a 24s time limit, however, we see that \yicesLS is able to solve around 700 benchmarks more than \yicesBase, 
 while, on the contrary, \hybridsmt loses around 450 benchmarks compared to Z3.
 \Cref{fig:cactus} shows that the point at which using local search pays off is much earlier for \yicesLS ($<10s$) than for \hybridsmt (just below $100s$).


%% file: results24s_vert.tex
\begin{table}[b!]
\caption{\label{fig:res24s} 
	Summary of results for  \nia benchmarks   with a timeout of 24s. 
}
	\setlength{\fboxsep}{0pt}
	\setlength{\tabcolsep}{0pt}
	\begin{scriptsize}		
		\begin{center}
  \begin{tabular}{c|>{\centering\arraybackslash}p{11ex}|>{\centering\arraybackslash}p{11ex}|>{\centering\arraybackslash}p{11ex}|>{\centering\arraybackslash}p{11ex}|>{\centering\arraybackslash}p{11ex}|>{\centering\arraybackslash}p{11ex}|>{\centering\arraybackslash}p{13ex}|>{\centering\arraybackslash}p{13ex}}  		
		\multirow{2}{*}{} &
		\rotatebox{\toprowtextangle}{\cvctool}  &
		\rotatebox{\toprowtextangle}{\hybridsmt}  &
		\rotatebox{\toprowtextangle}{\mathsat}  &
		\rotatebox{\toprowtextangle}{\yicesBase}  &
		\rotatebox{\toprowtextangle}{\yicesLS}  &
		\rotatebox{\toprowtextangle}{\zthree}  &
		\rotatebox{\toprowtextangle}{\stylePortfolio{\portfolioCellZ}}  &
		\rotatebox{\toprowtextangle}{\stylePortfolio{\portfolioCellHybrid}} 
	\\ \hline 
		VeryMax
		& \makecell{7799 \\ (5465) \\ (2334)}
		& \makecell{13555 \\ (\bestPartial{10427}) \\ (3128)}
		& \makecell{11233 \\ (7615) \\ (3618)}
		& \makecell{13695 \\ (9461) \\ (4234)}
		& \makecell{\bestTot{14269} \\ (10019) \\ (4250)}
		& \makecell{13975 \\ (9599) \\ (\bestPartial{4376})}
		& \stylePortfolio{\makecell{14848 \\ (10271) \\ (4577)}}
		& \stylePortfolio{\makecell{15428 \\ (11105) \\ (4323)}}\\ \hline 
		calypto
		& \makecell{171 \\ (79) \\ (92)}
		& \makecell{174 \\ (78) \\ (\bestPartial{96})}
		& \makecell{168 \\ (79) \\ (89)}
		& \makecell{174 \\ (79) \\ (95)}
		& \makecell{174 \\ (79) \\ (95)}
		& \makecell{\bestTot{176} \\ (\bestPartial{80}) \\ (\bestPartial{96})}
		& \stylePortfolio{\makecell{176 \\ (80) \\ (96)}}
		& \stylePortfolio{\makecell{176 \\ (80) \\ (96)}}\\ \hline 
		ezsmt
		& \makecell{\bestTot{8} \\ (\bestPartial{8}) \\ (\bestPartial{0})}
		& \makecell{\bestTot{8} \\ (\bestPartial{8}) \\ (\bestPartial{0})}
		& \makecell{\bestTot{8} \\ (\bestPartial{8}) \\ (\bestPartial{0})}
		& \makecell{\bestTot{8} \\ (\bestPartial{8}) \\ (\bestPartial{0})}
		& \makecell{\bestTot{8} \\ (\bestPartial{8}) \\ (\bestPartial{0})}
		& \makecell{\bestTot{8} \\ (\bestPartial{8}) \\ (\bestPartial{0})}
		& \stylePortfolio{\makecell{8 \\ (8) \\ (0)}}
		& \stylePortfolio{\makecell{8 \\ (8) \\ (0)}}\\ \hline 
		LassoRank
		& \makecell{97 \\ (\bestPartial{4}) \\ (93)}
		& \makecell{\bestTot{105} \\ (\bestPartial{4}) \\ (\bestPartial{101})}
		& \makecell{\bestTot{105} \\ (\bestPartial{4}) \\ (\bestPartial{101})}
		& \makecell{91 \\ (\bestPartial{4}) \\ (87)}
		& \makecell{96 \\ (\bestPartial{4}) \\ (92)}
		& \makecell{104 \\ (\bestPartial{4}) \\ (100)}
		& \stylePortfolio{\makecell{102 \\ (4) \\ (98)}}
		& \stylePortfolio{\makecell{104 \\ (4) \\ (100)}}\\ \hline 
		Dartagnan
		& \makecell{320 \\ (11) \\ (309)}
		& \makecell{\bestTot{354} \\ (10) \\ (\bestPartial{344})}
		& \makecell{327 \\ (\bestPartial{12}) \\ (315)}
		& \makecell{142 \\ (1) \\ (141)}
		& \makecell{87 \\ (0) \\ (87)}
		& \makecell{352 \\ (9) \\ (343)}
		& \stylePortfolio{\makecell{344 \\ (9) \\ (335)}}
		& \stylePortfolio{\makecell{347 \\ (9) \\ (338)}}\\ \hline 
		LCTES
		& \makecell{\bestTot{1} \\ (\bestPartial{0}) \\ (\bestPartial{1})}
		& \makecell{\bestTot{1} \\ (\bestPartial{0}) \\ (\bestPartial{1})}
		& \makecell{\bestTot{1} \\ (\bestPartial{0}) \\ (\bestPartial{1})}
		& \makecell{0 \\ (\bestPartial{0}) \\ (0)}
		& \makecell{0 \\ (\bestPartial{0}) \\ (0)}
		& \makecell{\bestTot{1} \\ (\bestPartial{0}) \\ (\bestPartial{1})}
		& \stylePortfolio{\makecell{1 \\ (0) \\ (1)}}
		& \stylePortfolio{\makecell{1 \\ (0) \\ (1)}}\\ \hline 
		MathProbl
		& \makecell{107 \\ (100) \\ (\bestPartial{7})}
		& \makecell{93 \\ (86) \\ (\bestPartial{7})}
		& \makecell{141 \\ (134) \\ (\bestPartial{7})}
		& \makecell{122 \\ (115) \\ (\bestPartial{7})}
		& \makecell{\bestTot{303} \\ (\bestPartial{296}) \\ (\bestPartial{7})}
		& \makecell{118 \\ (111) \\ (\bestPartial{7})}
		& \stylePortfolio{\makecell{300 \\ (293) \\ (7)}}
		& \stylePortfolio{\makecell{305 \\ (298) \\ (7)}}\\ \hline 
		leipgiz
		& \makecell{72 \\ (70) \\ (\bestPartial{2})}
		& \makecell{\bestTot{151} \\ (\bestPartial{150}) \\ (1)}
		& \makecell{114 \\ (112) \\ (\bestPartial{2})}
		& \makecell{102 \\ (101) \\ (1)}
		& \makecell{111 \\ (110) \\ (1)}
		& \makecell{120 \\ (119) \\ (1)}
		& \stylePortfolio{\makecell{119 \\ (118) \\ (1)}}
		& \stylePortfolio{\makecell{148 \\ (147) \\ (1)}}\\ \hline 
		UltAut23
		& \makecell{16 \\ (\bestPartial{8}) \\ (8)}
		& \makecell{10 \\ (7) \\ (3)}
		& \makecell{17 \\ (7) \\ (10)}
		& \makecell{7 \\ (7) \\ (0)}
		& \makecell{7 \\ (7) \\ (0)}
		& \makecell{\bestTot{21} \\ (\bestPartial{8}) \\ (\bestPartial{13})}
		& \stylePortfolio{\makecell{20 \\ (8) \\ (12)}}
		& \stylePortfolio{\makecell{9 \\ (7) \\ (2)}}\\ \hline 
		mcm
		& \makecell{9 \\ (9) \\ (\bestPartial{0})}
		& \makecell{\bestTot{56} \\ (\bestPartial{56}) \\ (\bestPartial{0})}
		& \makecell{3 \\ (3) \\ (\bestPartial{0})}
		& \makecell{6 \\ (6) \\ (\bestPartial{0})}
		& \makecell{6 \\ (6) \\ (\bestPartial{0})}
		& \makecell{5 \\ (5) \\ (\bestPartial{0})}
		& \stylePortfolio{\makecell{5 \\ (5) \\ (0)}}
		& \stylePortfolio{\makecell{46 \\ (46) \\ (0)}}\\ \hline 
		sqrtmodinv
		& \makecell{2 \\ (\bestPartial{0}) \\ (2)}
		& \makecell{7 \\ (\bestPartial{0}) \\ (7)}
		& \makecell{0 \\ (\bestPartial{0}) \\ (0)}
		& \makecell{0 \\ (\bestPartial{0}) \\ (0)}
		& \makecell{0 \\ (\bestPartial{0}) \\ (0)}
		& \makecell{\bestTot{17} \\ (\bestPartial{0}) \\ (\bestPartial{17})}
		& \stylePortfolio{\makecell{17 \\ (0) \\ (17)}}
		& \stylePortfolio{\makecell{6 \\ (0) \\ (6)}}\\ \hline 
		UltAut
		& \makecell{\bestTot{7} \\ (\bestPartial{0}) \\ (\bestPartial{7})}
		& \makecell{\bestTot{7} \\ (\bestPartial{0}) \\ (\bestPartial{7})}
		& \makecell{\bestTot{7} \\ (\bestPartial{0}) \\ (\bestPartial{7})}
		& \makecell{\bestTot{7} \\ (\bestPartial{0}) \\ (\bestPartial{7})}
		& \makecell{\bestTot{7} \\ (\bestPartial{0}) \\ (\bestPartial{7})}
		& \makecell{\bestTot{7} \\ (\bestPartial{0}) \\ (\bestPartial{7})}
		& \stylePortfolio{\makecell{7 \\ (0) \\ (7)}}
		& \stylePortfolio{\makecell{7 \\ (0) \\ (7)}}\\ \hline 
		AProVE
		& \makecell{1816 \\ (1251) \\ (565)}
		& \makecell{2212 \\ (1572) \\ (640)}
		& \makecell{2085 \\ (1556) \\ (529)}
		& \makecell{2328 \\ (1615) \\ (713)}
		& \makecell{\bestTot{2356} \\ (\bestPartial{1642}) \\ (\bestPartial{714})}
		& \makecell{2289 \\ (1612) \\ (677)}
		& \stylePortfolio{\makecell{2380 \\ (1657) \\ (723)}}
		& \stylePortfolio{\makecell{2371 \\ (1649) \\ (722)}}\\ \hline 
		UltLasso
		& \makecell{\bestTot{32} \\ (\bestPartial{6}) \\ (\bestPartial{26})}
		& \makecell{31 \\ (\bestPartial{6}) \\ (25)}
		& \makecell{\bestTot{32} \\ (\bestPartial{6}) \\ (\bestPartial{26})}
		& \makecell{\bestTot{32} \\ (\bestPartial{6}) \\ (\bestPartial{26})}
		& \makecell{\bestTot{32} \\ (\bestPartial{6}) \\ (\bestPartial{26})}
		& \makecell{\bestTot{32} \\ (\bestPartial{6}) \\ (\bestPartial{26})}
		& \stylePortfolio{\makecell{32 \\ (6) \\ (26)}}
		& \stylePortfolio{\makecell{32 \\ (6) \\ (26)}}
		\\ \hhline{=|=|=|=|=|=|=|=|=}
  	Total
  & \makecell{10457 \\ (7011) \\ (3446)}
  & \makecell{16764 \\ (\bestPartial{12404}) \\ (4360)}
  & \makecell{14241 \\ (9536) \\ (4705)}
  & \makecell{16714 \\ (11403) \\ (5311)}
  & \makecell{\bestTot{17456} \\ (12177) \\ (5279)}
  & \makecell{17225 \\ (11561) \\ (\bestPartial{5664})}
  & \stylePortfolio{\makecell{18359 \\ (12459) \\ (5900)}}
  & \stylePortfolio{\makecell{18988 \\ (13359) \\ (5629)}}
	\end{tabular}

\end{center}

\end{scriptsize}
\end{table}

%% file: results300s_vert.tex
\begin{table}[b!]
\caption{\label{fig:res300s} 
	Summary of results for  \nia benchmarks   with a timeout of 300s. 
}
	\setlength{\fboxsep}{0pt}
	\setlength{\tabcolsep}{0pt}
	\begin{scriptsize}		
		\begin{center}
  \begin{tabular}{c|>{\centering\arraybackslash}p{11ex}|>{\centering\arraybackslash}p{11ex}|>{\centering\arraybackslash}p{11ex}|>{\centering\arraybackslash}p{11ex}|>{\centering\arraybackslash}p{11ex}|>{\centering\arraybackslash}p{11ex}|>{\centering\arraybackslash}p{13ex}|>{\centering\arraybackslash}p{13ex}}  		
		\multirow{2}{*}{} &
		\rotatebox{\toprowtextangle}{\cvctool}  &
		\rotatebox{\toprowtextangle}{\hybridsmt}  &
		\rotatebox{\toprowtextangle}{\mathsat}  &
		\rotatebox{\toprowtextangle}{\yicesBase}  &
		\rotatebox{\toprowtextangle}{\yicesLS}  &
		\rotatebox{\toprowtextangle}{\zthree}  &
		\rotatebox{\toprowtextangle}{\stylePortfolio{\portfolioCellZ}}  &
		\rotatebox{\toprowtextangle}{\stylePortfolio{\portfolioCellHybrid}}  
		\\ \hline 
		VeryMax
		& \makecell{10489 \\ (7122) \\ (3367)}
		& \makecell{\bestTot{17060} \\ (\bestPartial{12093}) \\ (4967)}
		& \makecell{13651 \\ (9509) \\ (4142)}
		& \makecell{14549 \\ (10164) \\ (4385)}
		& \makecell{15160 \\ (10719) \\ (4441)}
		& \makecell{16304 \\ (11044) \\ (\bestPartial{5260})}
		& \stylePortfolio{\makecell{16698 \\ (11366) \\ (5332)}}
		& \stylePortfolio{\makecell{17303 \\ (12156) \\ (5147)}}\\ \hline 
		calypto
		& \makecell{173 \\ (79) \\ (94)}
		& \makecell{175 \\ (78) \\ (\bestPartial{97})}
		& \makecell{169 \\ (79) \\ (90)}
		& \makecell{174 \\ (79) \\ (95)}
		& \makecell{175 \\ (79) \\ (96)}
		& \makecell{\bestTot{177} \\ (\bestPartial{80}) \\ (\bestPartial{97})}
		& \stylePortfolio{\makecell{177 \\ (80) \\ (97)}}
		& \stylePortfolio{\makecell{177 \\ (80) \\ (97)}}\\ \hline 
		ezsmt
		& \makecell{\bestTot{8} \\ (\bestPartial{8}) \\ (\bestPartial{0})}
		& \makecell{\bestTot{8} \\ (\bestPartial{8}) \\ (\bestPartial{0})}
		& \makecell{\bestTot{8} \\ (\bestPartial{8}) \\ (\bestPartial{0})}
		& \makecell{\bestTot{8} \\ (\bestPartial{8}) \\ (\bestPartial{0})}
		& \makecell{\bestTot{8} \\ (\bestPartial{8}) \\ (\bestPartial{0})}
		& \makecell{\bestTot{8} \\ (\bestPartial{8}) \\ (\bestPartial{0})}
		& \stylePortfolio{\makecell{8 \\ (8) \\ (0)}}
		& \stylePortfolio{\makecell{8 \\ (8) \\ (0)}}\\ \hline 
		LassoRank
		& \makecell{98 \\ (\bestPartial{4}) \\ (94)}
		& \makecell{\bestTot{106} \\ (\bestPartial{4}) \\ (\bestPartial{102})}
		& \makecell{105 \\ (\bestPartial{4}) \\ (101)}
		& \makecell{93 \\ (\bestPartial{4}) \\ (89)}
		& \makecell{97 \\ (\bestPartial{4}) \\ (93)}
		& \makecell{\bestTot{106} \\ (\bestPartial{4}) \\ (\bestPartial{102})}
		& \stylePortfolio{\makecell{104 \\ (4) \\ (100)}}
		& \stylePortfolio{\makecell{105 \\ (4) \\ (101)}}\\ \hline 
		Dartagnan
		& \makecell{350 \\ (17) \\ (333)}
		& \makecell{\bestTot{369} \\ (14) \\ (\bestPartial{355})}
		& \makecell{347 \\ (\bestPartial{18}) \\ (329)}
		& \makecell{311 \\ (7) \\ (304)}
		& \makecell{288 \\ (3) \\ (285)}
		& \makecell{368 \\ (14) \\ (354)}
		& \stylePortfolio{\makecell{363 \\ (13) \\ (350)}}
		& \stylePortfolio{\makecell{367 \\ (13) \\ (354)}}\\ \hline 
		LCTES
		& \makecell{1 \\ (\bestPartial{0}) \\ (1)}
		& \makecell{\bestTot{2} \\ (\bestPartial{0}) \\ (\bestPartial{2})}
		& \makecell{1 \\ (\bestPartial{0}) \\ (1)}
		& \makecell{0 \\ (\bestPartial{0}) \\ (0)}
		& \makecell{0 \\ (\bestPartial{0}) \\ (0)}
		& \makecell{\bestTot{2} \\ (\bestPartial{0}) \\ (\bestPartial{2})}
		& \stylePortfolio{\makecell{1 \\ (0) \\ (1)}}
		& \stylePortfolio{\makecell{1 \\ (0) \\ (1)}}\\ \hline 
		MathProbl
		& \makecell{177 \\ (170) \\ (\bestPartial{7})}
		& \makecell{107 \\ (100) \\ (\bestPartial{7})}
		& \makecell{183 \\ (176) \\ (\bestPartial{7})}
		& \makecell{124 \\ (117) \\ (\bestPartial{7})}
		& \makecell{\bestTot{311} \\ (\bestPartial{304}) \\ (\bestPartial{7})}
		& \makecell{118 \\ (111) \\ (\bestPartial{7})}
		& \stylePortfolio{\makecell{314 \\ (307) \\ (7)}}
		& \stylePortfolio{\makecell{327 \\ (320) \\ (7)}}\\ \hline 
		leipgiz
		& \makecell{89 \\ (87) \\ (\bestPartial{2})}
		& \makecell{\bestTot{156} \\ (\bestPartial{155}) \\ (1)}
		& \makecell{126 \\ (124) \\ (\bestPartial{2})}
		& \makecell{104 \\ (103) \\ (1)}
		& \makecell{111 \\ (110) \\ (1)}
		& \makecell{135 \\ (134) \\ (1)}
		& \stylePortfolio{\makecell{134 \\ (133) \\ (1)}}
		& \stylePortfolio{\makecell{154 \\ (153) \\ (1)}}\\ \hline 
		UltAut23
		& \makecell{16 \\ (\bestPartial{8}) \\ (0)}
		& \makecell{10 \\ (7) \\ (3)}
		& \makecell{17 \\ (7) \\ (10)}
		& \makecell{7 \\ (7) \\ (0)}
		& \makecell{7 \\ (7) \\ (0)}
		& \makecell{\bestTot{21} \\ (\bestPartial{8}) \\ (\bestPartial{13})}
		& \stylePortfolio{\makecell{21 \\ (8) \\ (13)}}
		& \stylePortfolio{\makecell{10 \\ (7) \\ (3)}}\\ \hline 
		mcm
		& \makecell{17 \\ (17) \\ (\bestPartial{0})}
		& \makecell{\bestTot{69} \\ (\bestPartial{69}) \\ (\bestPartial{0})}
		& \makecell{10 \\ (10) \\ (\bestPartial{0})}
		& \makecell{9 \\ (9) \\ (\bestPartial{0})}
		& \makecell{9 \\ (9) \\ (\bestPartial{0})}
		& \makecell{10 \\ (10) \\ (\bestPartial{0})}
		& \stylePortfolio{\makecell{11 \\ (11) \\ (0)}}
		& \stylePortfolio{\makecell{64 \\ (64) \\ (0)}}\\ \hline 
		sqrtmodinv
		& \makecell{2 \\ (\bestPartial{0}) \\ (2)}
		& \makecell{10 \\ (\bestPartial{0}) \\ (10)}
		& \makecell{0 \\ (\bestPartial{0}) \\ (0)}
		& \makecell{0 \\ (\bestPartial{0}) \\ (0)}
		& \makecell{0 \\ (\bestPartial{0}) \\ (0)}
		& \makecell{\bestTot{17} \\ (\bestPartial{0}) \\ (\bestPartial{17})}
		& \stylePortfolio{\makecell{17 \\ (0) \\ (17)}}
		& \stylePortfolio{\makecell{8 \\ (0) \\ (8)}}\\ \hline 
		UltAut
		& \makecell{\bestTot{7} \\ (\bestPartial{0}) \\ (\bestPartial{7})}
		& \makecell{\bestTot{7} \\ (\bestPartial{0}) \\ (\bestPartial{7})}
		& \makecell{\bestTot{7} \\ (\bestPartial{0}) \\ (\bestPartial{7})}
		& \makecell{\bestTot{7} \\ (\bestPartial{0}) \\ (\bestPartial{7})}
		& \makecell{\bestTot{7} \\ (\bestPartial{0}) \\ (\bestPartial{7})}
		& \makecell{\bestTot{7} \\ (\bestPartial{0}) \\ (\bestPartial{7})}
		& \stylePortfolio{\makecell{7 \\ (0) \\ (7)}}
		& \stylePortfolio{\makecell{7 \\ (0) \\ (7)}}\\ \hline 
		AProVE
		& \makecell{1939 \\ (1330) \\ (609)}
		& \makecell{2352 \\ (1640) \\ (685)}
		& \makecell{2180 \\ (1622) \\ (558)}
		& \makecell{2337 \\ (1621) \\ (716)}
		& \makecell{\bestTot{2367} \\ (\bestPartial{1647}) \\ (\bestPartial{720})}
		& \makecell{2339 \\ (1640) \\ (699)}
		& \stylePortfolio{\makecell{2394 \\ (1661) \\ (733)}}
		& \stylePortfolio{\makecell{2387 \\ (1656) \\ (731)}}\\ \hline 
		UltLasso
		& \makecell{\bestTot{32} \\ (\bestPartial{6}) \\ (\bestPartial{26})}
		& \makecell{31 \\ (\bestPartial{6}) \\ (25)}
		& \makecell{\bestTot{32} \\ (\bestPartial{6}) \\ (\bestPartial{26})}
		& \makecell{\bestTot{32} \\ (\bestPartial{6}) \\ (\bestPartial{26})}
		& \makecell{\bestTot{32} \\ (\bestPartial{6}) \\ (\bestPartial{26})}
		& \makecell{\bestTot{32} \\ (\bestPartial{6}) \\ (\bestPartial{26})}
		& \stylePortfolio{\makecell{32 \\ (6) \\ (26)}}
		& \stylePortfolio{\makecell{32 \\ (6) \\ (26)}}
		\\ \hhline{=|=|=|=|=|=|=|=|=}
		Total
		& \makecell{13398 \\ (8848) \\ (4550)}
		& \makecell{\bestTot{20435} \\ (\bestPartial{14174}) \\ (6261)}
		& \makecell{16836 \\ (11563) \\ (5273)}
		& \makecell{17755 \\ (12125) \\ (5630)}
		& \makecell{18572 \\ (12896) \\ (5676)}
		& \makecell{19644 \\ (13059) \\ (\bestPartial{6585})}
		& \stylePortfolio{\makecell{20281 \\ (13597) \\ (6684)}}
		& \stylePortfolio{\makecell{20950 \\ (14467) \\ (6483)}}
	\end{tabular}

\end{center}

\end{scriptsize}
\end{table}

%% file: cactus_plots.tex
\begin{figure}[b!]
	\centering
		\includegraphics[trim = 0cm .8cm 0cm 1.4cm, clip, scale=0.8]{"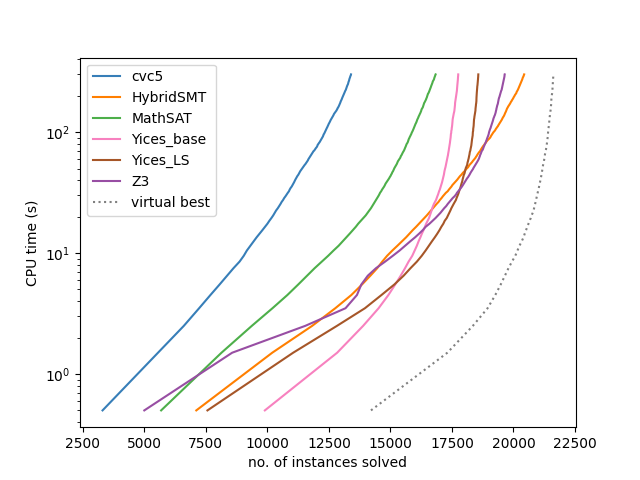"}
\caption{Plots showing the number of instances solved (x axis) within given time in seconds (y axis) in log scale.}
	\label{fig:cactus}
\end{figure}

%% file: scatter.tex
\begin{figure}[b!]
\def\sclen{0.496\textwidth}
\def\scscale{0.52}
\centering
	\begin{subfigure}{\sclen}
		\includegraphics[trim = 1.3cm 0cm 0cm 1.4cm, clip, scale=\scscale]{"RES/scatter\_plots/nia.922ea904/grayscale\_friendly/vs\_cvc5.png"}
	\end{subfigure}
	\begin{subfigure}{\sclen}
	\includegraphics[trim = 1.2cm 0cm 0cm 1.4cm, clip, scale=\scscale]{"RES/scatter\_plots/nia.922ea904/grayscale\_friendly/vs\_HybridSMT.png"}
	\end{subfigure}

	\begin{subfigure}{\sclen}
		\includegraphics[trim = 1.2cm 0cm 0cm 1.4cm, clip, scale=\scscale]{"RES/scatter\_plots/nia.922ea904/grayscale\_friendly/vs\_MathSAT.png"}
	\end{subfigure}	
	\begin{subfigure}{\sclen}
	\includegraphics[trim = 1.2cm 0cm 0cm 1.4cm, clip, scale=\scscale]{"RES/scatter\_plots/nia.922ea904/grayscale\_friendly/vs\_YicesBase.png"}
	\end{subfigure}	
\begin{subfigure}{\sclen}
	\includegraphics[trim = 1.2cm 0cm 0cm 1.4cm, clip, scale=\scscale]{"RES/scatter\_plots/nia.922ea904/grayscale\_friendly/vs\_Z3.png"}
\end{subfigure}
	\caption{Scatter plots comparing \yicesLS to \cvctool, \hybridsmt, \mathsat, \yicesBase, and \zthree, respectively;
		 on  sat.\ (orange) and unsat.\ (blue) instances.}
	\label{fig:scatter}
\end{figure}

%% file: relwork.tex
\section{Related work}
\label{sec:relwork}
 
In propositional SAT solving, local search techniques have been successfully used to solve difficult satisfiable problems~\cite{LSHandbookSAT} as well as unsatisfiable instances~\cite{LSUnsat}. 
Recently, their tight integration in the propositional CDCL framework has been shown to improve performance~\cite{deepCombLsSat,satCdclLStargetphase} and are now considered a key component of \sota SAT solvers.
In the context of SMT, on the other hand, the adoption of LS is a lot less widespread. 

In~\cite{SLSWalkSAT}, the LS-based SAT solver WalkSAT has been used in combination with a theory solver as an alternative to the classic \cdclt approach; however,
 the use of  local search remained limited to the Boolean level.
For the theory of bit-vectors, the idea of Boolean flips in SAT solving has been transposed to the bit level by introducing \emph{bit-flips}~\cite{SLSBVflips}, possibly augmented with propagations~\cite{SLSBVProp}.

The adoption of LS for arithmetic theories is more recent.
For the theories of Linear Integer Arithmetic (\lia)~\cite{LSLIA} and Multi-linear Real Arithmetic~\cite{LSMultilinRA} 
  a \emph{critical move} operation is used to change the value of a variable that appears in a literal violated by the current assignment in order to make the literal satisfied.
 To deal with the nonlinear arithmetic constraints,
 the \emph{cell-jumping} technique is used, which first isolates the roots of a falsified polynomial \wrt to a variable (by fixing the value of the other variables), thus 
 decomposing the real space into finitely many intervals (\emph{cells}, in the CAD~\cite{collins} terminology), 
 and then tries to satisfy the polynomial by changing the value of the variable by {jumping} around these cells.
 This technique has been implemented in LocalSMT for \nia~\cite{LocalSearchIA}, and as a tool in Maple~\cite{LocalSearch4SatPolFormulas} and on top of Z3~\cite{EfficientLocalSearch} for \nra.
 %
 %
%

Local search has also  been used as a sub-routine for \emph{global search} techniques, as in the case of 
floating points ~\cite{xsat},
and of \nra possibly augmented with transcendental functions (\nta)~\cite{ATVApaper,LippariniRatschanJAR,LippariniThesis}.
In these works, numerical optimization algorithms, \eg the gradient-descent, are used to find local minima, while stochastic jumping is used to move away from a local minimum in order to explore other regions in search for a global minimum.

All the methods discussed so far for arithmetic theories are only able to prove satisfiability; 
if they fail, then all the knowledge that has been acquired by 
the search 
 is lost.
%
 HybridSMT~\cite{localsearchCDCLT} addresses this issue, for the case of \nia,
by integrating LocalSMT within Z3's \cdclt. In particular, LocalSMT takes as input a subformula corresponding to a Boolean skeleton solution, and, if it does not find an integer solution for the subformula, it returns	the best assignment found and the conflict frequency for atoms. This information is used to improve phase selection (\ie Boolean assignments) and variable ordering.
 Although both HybridSMT and our method share the idea of integrating LS within a reasoning calculus (\cdclt and \mcsat, respectively), there are some substantial difference.
 First, in HybridSMT, LS takes into account complete Boolean variable assignments. In our framework, LS can take as input both Boolean and theory variable assignments, either partial or complete. 
 Additionally, while in HybridSMT LS can only suggest assignments for (and ordering of) Boolean literals, we extend that to theory variables as well.
 Moreover, there is a theory-specific difference in our approach. LocalSMT relies on cell-jumps, which require to perform potentially very expensive root isolation sub-routines 
 at every step. 
 In contrast, our method uses fs-jumps that rely on feasibility intervals already maintained by the theory plugin in the MCSat framework. This eliminates the need for additional computation and can be viewed as a lazy version of cells, progressively refined on-demand. 
  Furthermore, we pair fs-jumps with hill-climbing to move inside feasible intervals.

 Most \sota solvers do not use local search for \nia problems.
 Bit-blasting~\cite{bitblasting} aims at proving satisfiability by iteratively imposing bounds on the variables 
 and then encoding the obtained sub-formula into an equi-satisfiable Boolean formula, which is then handled by a SAT solver. 
 %
 In the branch-and-bound approach~\cite{branchBoundKrem,branchBoundJovanovic} the integer domain is relaxed by allowing variables to range over real numbers.
 %
 Incremental Linearization~\cite{incrlinNIA,incrlin}  leverages decision procedures for \lia by  abstracting non-linear multiplications with uninterpreted functions and then
 incrementally axiomatize them.

%% file: conclusion.tex
\section{Conclusion}
\label{sec:conclusion}

In this work, we have introduced a theory-independent framework for
integrating local search into the \mcsat calculus. By combining local
search intuition with \mcsat reasoning capabilities, our approach
leverages logic-to-optimization formalization to provide guidance to
the core \mcsat solver. Specifically, we addressed the theory of
nonlinear integer arithmetic by proposing a local search procedure
based on feasibility-set jumping and hill-climbing.

We implemented our approach in the \yices SMT solver, and empirically
demonstrated its improvements for both satisfiable and unsatisfiable
instances.  Our results show that the new \yices solver with local
search compares favorably and often outperforms other SMT solvers; in
particular, it manages to solve a significant amount of benchmarks not
solved by other \sota tools.  Moreover, our results show that the new
\yices is able to reach solutions or proofs more efficiently, compared
to the baseline \yices, in terms of the number of decisions and
conflicts.
Our findings indicate that this new approach complements existing
techniques used by other solvers, as evident in the experimental results
(see virtual best and
portfolio solvers in the plots and tables).

In the future, we aim to extend our approach to other theories such as
finite fields and bit-vectors, and conduct more comprehensive
experimental evaluations. Additionally, we plan to integrate this
approach with different caching schemes, including value and target
caches, along with periodic recaching in \mcsat~\cite{cav25}.

%% file: main.bbl
\begin{thebibliography}{10}

\bibitem{cvc5}
Haniel Barbosa, Clark~W. Barrett, Martin Brain, Gereon Kremer, Hanna Lachnitt,
  Makai Mann, Abdalrhman Mohamed, Mudathir Mohamed, Aina Niemetz, Andres
  N{\"{o}}tzli, Alex Ozdemir, Mathias Preiner, Andrew Reynolds, Ying Sheng,
  Cesare Tinelli, and Yoni Zohar.
\newblock cvc5: {A} versatile and industrial-strength {SMT} solver.
\newblock In Dana Fisman and Grigore Rosu, editors, {\em Tools and Algorithms
  for the Construction and Analysis of Systems - 28th International Conference,
  {TACAS} 2022, Held as Part of the European Joint Conferences on Theory and
  Practice of Software, {ETAPS} 2022, Munich, Germany, April 2-7, 2022,
  Proceedings, Part {I}}, volume 13243 of {\em Lecture Notes in Computer
  Science}, pages 415--442. Springer, 2022.

\bibitem{SMTLIB}
Clark Barrett, Pascal Fontaine, and Cesare Tinelli.
\newblock {The Satisfiability Modulo Theories Library (SMT-LIB)}.
\newblock {\tt www.SMT-LIB.org}, 2016.

\bibitem{smt}
Clark~W. Barrett, Roberto Sebastiani, Sanjit~A. Seshia, and Cesare Tinelli.
\newblock {Satisfiability Modulo Theories}.
\newblock In Armin Biere, Marijn Heule, Hans van Maaren, and Toby Walsh,
  editors, {\em Handbook of Satisfiability - Second Edition}, volume 336 of
  {\em Frontiers in Artificial Intelligence and Applications}, pages
  1267--1329. {IOS} Press, 2021.

\bibitem{z3ArithSolving}
Nikolaj Bj{\o}rner and Lev Nachmanson.
\newblock Arithmetic solving in z3.
\newblock In Arie Gurfinkel and Vijay Ganesh, editors, {\em Computer Aided
  Verification}, pages 26--41, Cham, 2024. Springer Nature Switzerland.

\bibitem{LSLIA}
Shaowei Cai, Bohan Li, and Xindi Zhang.
\newblock {Local Search for SMT on Linear Integer Arithmetic}.
\newblock In Sharon Shoham and Yakir Vizel, editors, {\em Computer Aided
  Verification}, pages 227--248, Cham, 2022. Springer International Publishing.

\bibitem{LocalSearchIA}
Shaowei Cai, Bohan Li, and Xindi Zhang.
\newblock Local search for satisfiability modulo integer arithmetic theories.
\newblock {\em ACM Trans. Comput. Logic}, 24(4), July 2023.

\bibitem{deepCombLsSat}
Shaowei Cai and Xindi Zhang.
\newblock Deep cooperation of {CDCL} and local search for {SAT}.
\newblock In Chu-Min Li and Felip Many{\`a}, editors, {\em Theory and
  Applications of Satisfiability Testing -- SAT 2021}, pages 64--81, Cham,
  2021. Springer International Publishing.

\bibitem{satCdclLStargetphase}
Shaowei Cai, Xindi Zhang, Mathias Fleury, and Armin Biere.
\newblock Better decision heuristics in {CDCL} through local search and target
  phases.
\newblock {\em J. Artif. Int. Res.}, 74, September 2022.

\bibitem{incrlinNIA}
Alessandro Cimatti, Alberto Griggio, Ahmed Irfan, Marco Roveri, and Roberto
  Sebastiani.
\newblock Experimenting on solving nonlinear integer arithmetic with
  incremental linearization.
\newblock In Olaf Beyersdorff and Christoph~M. Wintersteiger, editors, {\em
  Theory and Applications of Satisfiability Testing -- SAT 2018}, pages
  383--398, Cham, 2018. Springer International Publishing.

\bibitem{incrlin}
Alessandro Cimatti, Alberto Griggio, Ahmed Irfan, Marco Roveri, and Roberto
  Sebastiani.
\newblock Incremental linearization for satisfiability and verification modulo
  nonlinear arithmetic and transcendental functions.
\newblock {\em ACM Trans. Comput. Logic}, 19(3), aug 2018.

\bibitem{mathsat5}
Alessandro Cimatti, Alberto Griggio, Bastiaan Schaafsma, and Roberto
  Sebastiani.
\newblock {The {MathSAT5} {SMT} Solver}.
\newblock In Nir Piterman and Scott Smolka, editors, {\em Proceedings of
  TACAS}, volume 7795 of {\em LNCS}. Springer, 2013.

\bibitem{collins}
George~E. Collins.
\newblock Quantifier elimination for real closed fields by cylindrical
  algebraic decompostion.
\newblock In H.~Brakhage, editor, {\em Automata Theory and Formal Languages},
  pages 134--183, Berlin, Heidelberg, 1975. Springer Berlin Heidelberg.

\bibitem{SMTRAT}
Florian Corzilius, Gereon Kremer, Sebastian Junges, Stefan Schupp, and Erika
  Ábrahám.
\newblock {SMT-RAT}: An open source {C++} toolbox for strategic and parallel
  {SMT} solving.
\newblock In {\em SAT}, 09 2015.

\bibitem{z3}
Leonardo De~Moura and Nikolaj Bj\o{}rner.
\newblock Z3: An efficient {SMT} solver.
\newblock In {\em Proceedings of the Theory and Practice of Software, 14th
  International Conference on Tools and Algorithms for the Construction and
  Analysis of Systems}, TACAS'08/ETAPS'08, page 337–340, Berlin, Heidelberg,
  2008. Springer-Verlag.

\bibitem{mcsat}
Leonardo de~Moura and Dejan Jovanovic.
\newblock A model-constructing satisfiability calculus.
\newblock In Roberto Giacobazzi, Josh Berdine, and Isabella Mastroeni, editors,
  {\em Intl. Conference on Verification, Model Checking, and Abstract
  Interpretation {(VMCAI)}}, volume 7737 of {\em LNCS}, pages 1--12. Springer,
  2013.

\bibitem{DBLP:journals/cacm/MouraB11}
Leonardo~Mendon{\c{c}}a de~Moura and Nikolaj~S. Bj{\o}rner.
\newblock Satisfiability modulo theories: introduction and applications.
\newblock {\em Commun. {ACM}}, 54(9):69--77, 2011.

\bibitem{yices}
Bruno Dutertre.
\newblock Yices 2.2.
\newblock In Armin Biere and Roderick Bloem, editors, {\em Computer-Aided
  Verification (CAV'2014)}, volume 8559 of {\em Lecture Notes in Computer
  Science}, pages 737--744. Springer, July 2014.

\bibitem{SLSBVflips}
Andreas Fröhlich, Armin Biere, Christoph Wintersteiger, and Youssef Hamadi.
\newblock Stochastic local search for {Satisfiability Modulo Theories}.
\newblock {\em Proceedings of the AAAI Conference on Artificial Intelligence},
  29(1), Feb. 2015.

\bibitem{xsat}
Zhoulai Fu and Zhendong Su.
\newblock {XSat}: {A} fast floating-point satisfiability solver.
\newblock In {\em {CAV}}, volume 9780 of {\em Lecture Notes in Computer
  Science}, pages 187--209. Springer, 2016.

\bibitem{bitblasting}
Carsten Fuhs, J{\"u}rgen Giesl, Aart Middeldorp, Peter Schneider-Kamp, Ren{\'e}
  Thiemann, and Harald Zankl.
\newblock {SAT} solving for termination analysis with polynomial
  interpretations.
\newblock In Jo{\~a}o Marques-Silva and Karem~A. Sakallah, editors, {\em Theory
  and Applications of Satisfiability Testing -- SAT 2007}, pages 340--354,
  Berlin, Heidelberg, 2007. Springer Berlin Heidelberg.

\bibitem{mcsatBV}
St{\'{e}}phane Graham{-}Lengrand, Dejan Jovanovic, and Bruno Dutertre.
\newblock Solving bitvectors with {MCSAT:} explanations from bits and pieces.
\newblock In Nicolas Peltier and Viorica Sofronie{-}Stokkermans, editors, {\em
  Intl. Joint Conf. on Automated Reasoning {(IJCAR)}, Part {I}}, volume 12166
  of {\em LNCS}, pages 103--121. Springer, 2020.

\bibitem{SLSWalkSAT}
Alberto Griggio, Quoc-Sang Phan, Roberto Sebastiani, and Silvia Tomasi.
\newblock Stochastic local search for {SMT}: Combining theory solvers with
  {WalkSAT}.
\newblock In Cesare Tinelli and Viorica Sofronie-Stokkermans, editors, {\em
  Frontiers of Combining Systems}, pages 163--178, Berlin, Heidelberg, 2011.
  Springer Berlin Heidelberg.

\bibitem{cav25}
Thomas Hader, Ahmed Irfan, and St\'ephane Graham-Lengrand.
\newblock Decision heuristics in {MCSat} \textit{(to appear)}.
\newblock In {\em Computer Aided Verification}, 2025.

\bibitem{mcsatFF2}
Thomas Hader, Daniela Kaufmann, Ahmed Irfan, St{\'{e}}phane Graham{-}Lengrand,
  and Laura Kov{\'{a}}cs.
\newblock {MCSat}-based finite field reasoning in the {Yices2} {SMT} solver
  (short paper).
\newblock In {\em {IJCAR} {(1)}}, volume 14739 of {\em Lecture Notes in
  Computer Science}, pages 386--395. Springer, 2024.

\bibitem{mcsatFF1}
Thomas Hader, Daniela Kaufmann, and Laura Kov{\'{a}}cs.
\newblock {SMT} solving over finite field arithmetic.
\newblock In {\em {LPAR}}, volume~94 of {\em EPiC Series in Computing}, pages
  238--256. EasyChair, 2023.

\bibitem{HillClimbing}
Leticia Hernando, Alexander Mendiburu, and Jose Lozano.
\newblock Hill-climbing algorithm: Let's go for a walk before finding the
  optimum.
\newblock pages 1--7, 07 2018.

\bibitem{mcsatArrays}
Ahmed Irfan and St{\'{e}}phane Graham{-}Lengrand.
\newblock Arrays reasoning in {MCSat}.
\newblock In {\em SMT@CAV}, volume 3725 of {\em {CEUR} Workshop Proceedings},
  pages 24--35. CEUR-WS.org, 2024.

\bibitem{branchBoundJovanovic}
Dejan Jovanovi{\'{c}}.
\newblock Solving nonlinear integer arithmetic with {MCSAT}.
\newblock In Ahmed Bouajjani and David Monniaux, editors, {\em Verification,
  Model Checking, and Abstract Interpretation}, pages 330--346, Cham, 2017.
  Springer International Publishing.

\bibitem{mcsat2}
Dejan Jovanovic, Clark Barrett, and Leonardo de~Moura.
\newblock The design and implementation of the model constructing
  satisfiability calculus.
\newblock In {\em Intl. Conf on Formal Methods in Computer-Aided Design
  {(FMCAD)}}, pages 173--180. IEEE, 2013.

\bibitem{solving_nonlin}
Dejan Jovanovi\'{c} and Leonardo de~Moura.
\newblock Solving non-linear arithmetic.
\newblock {\em ACM Commun. Comput. Algebra}, 46(3/4):104–105, jan 2013.

\bibitem{LSHandbookSAT}
Henry~A. Kautz, Ashish Sabharwal, and Bart Selman.
\newblock Incomplete algorithms.
\newblock In Armin Biere, Marijn Heule, Hans van Maaren, and Toby Walsh,
  editors, {\em Handbook of Satisfiability - Second Edition}, volume 336 of
  {\em Frontiers in Artificial Intelligence and Applications}, pages 213--232.
  {IOS} Press, 2021.

\bibitem{branchBoundKrem}
Gereon Kremer, Florian Corzilius, and Erika {\'{A}}brah{\'{a}}m.
\newblock A generalised branch-and-bound approach and its application in {SAT}
  modulo nonlinear integer arithmetic.
\newblock In Vladimir~P. Gerdt, Wolfram Koepf, Werner~M. Seiler, and Evgenii~V.
  Vorozhtsov, editors, {\em Computer Algebra in Scientific Computing - 18th
  International Workshop, {CASC} 2016, Bucharest, Romania, September 19-23,
  2016, Proceedings}, volume 9890 of {\em Lecture Notes in Computer Science},
  pages 315--335. Springer, 2016.

\bibitem{SMTRATmodular}
Gereon Kremer and Erika Ábrahám.
\newblock Modular strategic {SMT} solving with {SMT-RAT}.
\newblock {\em Acta Universitatis Sapientiae, Informatica}, 10(1):5--25, 2018.

\bibitem{LSMultilinRA}
Bohan Li and Shaowei Cai.
\newblock Local search for {SMT} on linear and multi-linear real arithmetic.
\newblock In {\em 2023 Formal Methods in Computer-Aided Design (FMCAD)}, pages
  1--10, 2023.

\bibitem{LocalSearch4SatPolFormulas}
Haokun Li, Bican Xia, and Tianqi Zhao.
\newblock Local search for solving satisfiability of polynomial formulas.
\newblock In Constantin Enea and Akash Lal, editors, {\em Computer Aided
  Verification}, pages 87--109, Cham, 2023. Springer Nature Switzerland.

\bibitem{LippariniThesis}
Enrico Lipparini.
\newblock {\em Satisfiability modulo Nonlinear Arithmetic and Transcendental
  Functions via Numerical and Topological methods}.
\newblock PhD thesis, 2024.

\bibitem{ATVApaper}
Enrico Lipparini, Alessandro Cimatti, Alberto Griggio, and Roberto Sebastiani.
\newblock Handling polynomial and transcendental functions in {SMT}
  via unconstrained optimisation and topological degree test.
\newblock In Ahmed Bouajjani, Luk{\'a}{\v{s}} Hol{\'i}k, and Zhilin Wu,
  editors, {\em Automated Technology for Verification and Analysis}, pages
  137--153, Cham, 2022. Springer International Publishing.

\bibitem{Lipparini:23}
Enrico Lipparini and Stefan Ratschan.
\newblock Satisfiability of non-linear transcendental arithmetic
  as a certificate search problem.
\newblock In Kristin~Yvonne Rozier and Swarat Chaudhuri, editors, {\em NASA
  Formal Methods}, pages 472--488, Cham, 2023. Springer Nature Switzerland.

\bibitem{LippariniRatschanJAR}
Enrico Lipparini and Stefan Ratschan.
\newblock Satisfiability of non-linear transcendental arithmetic as a
  certificate search problem.
\newblock {\em J. Autom. Reason.}, 69(1), January 2025.

\bibitem{Matiyasevich}
Yuri~V. Matiyasevich.
\newblock {\em Hilbert's tenth problem}.
\newblock MIT Press, Cambridge, MA, USA, 1993.

\bibitem{SLSBVProp}
Aina Niemetz, Mathias Preiner, and Armin Biere.
\newblock Propagation based local search for bit-precise reasoning.
\newblock {\em Formal Methods in System Design}, 51(3):608--636, Dec 2017.

\bibitem{dpllt}
Robert Nieuwenhuis, Albert Oliveras, and Cesare Tinelli.
\newblock Solving {SAT} and {SAT} modulo theories: From an abstract
  {Davis--Putnam--Logemann--Loveland} procedure to {DPLL(T)}.
\newblock {\em J. ACM}, 53(6):937–977, November 2006.

\bibitem{pipatsrisawat2007lightweight}
Knot Pipatsrisawat and Adnan Darwiche.
\newblock A lightweight component caching scheme for satisfiability solvers.
\newblock In {\em Theory and Applications of Satisfiability Testing--SAT 2007:
  10th International Conference, Lisbon, Portugal, May 28-31, 2007. Proceedings
  10}, pages 294--299. Springer, 2007.

\bibitem{preiner_2024_11061097}
Mathias Preiner, Hans-Jörg Schurr, Clark Barrett, Pascal Fontaine, Aina
  Niemetz, and Cesare Tinelli.
\newblock {SMT-LIB} release 2024 (non-incremental benchmarks), April 2024.

\bibitem{LSUnsat}
Steven Prestwich and In{\^e}s Lynce.
\newblock Local search for unsatisfiability.
\newblock In Armin Biere and Carla~P. Gomes, editors, {\em Theory and
  Applications of Satisfiability Testing - SAT 2006}, pages 283--296, Berlin,
  Heidelberg, 2006. Springer Berlin Heidelberg.

\bibitem{EfficientLocalSearch}
Zhonghan Wang, Bohua Zhan, Bohan Li, and Shaowei Cai.
\newblock Efficient local search for nonlinear real arithmetic.
\newblock In Rayna Dimitrova, Ori Lahav, and Sebastian Wolff, editors, {\em
  Verification, Model Checking, and Abstract Interpretation}, pages 326--349,
  Cham, 2024. Springer Nature Switzerland.

\bibitem{SMTComp}
Tjark Weber, Sylvain Conchon, David D{\'{e}}harbe, Matthias Heizmann, Aina
  Niemetz, and Giles Reger.
\newblock The {SMT} competition 2015-2018.
\newblock {\em J. Satisf. Boolean Model. Comput.}, 11(1):221--259, 2019.

\bibitem{localsearchCDCLT}
Xindi Zhang, Bohan Li, and Shaowei Cai.
\newblock Deep combination of {CDCL(T)} and local search for satisfiability
  modulo non-linear integer arithmetic theory.
\newblock In {\em Proceedings of the IEEE/ACM 46th International Conference on
  Software Engineering}, ICSE '24, New York, NY, USA, 2024. Association for
  Computing Machinery.

\end{thebibliography}
